\documentclass[useAMS,usenatbib,usegraphicx]{mn2e}

\title[Testing $z\ga7$ Photometric Ly$\alpha$ Luminosity Functions]{Subaru FOCAS Survey of $z=7$--$7.1$ Ly$\alpha$ Emitters: A Test for $z\ga7$ Ly$\alpha$ Photometric Luminosity Functions}
\author[Kazuaki Ota and Masanori Iye]{Kazuaki Ota$^{1}$\thanks{E-mail:
otakz@kusastro.kyoto-u.ac.jp} and Masanori Iye$^{2,3}$\\
$^{1}$Department of Astronomy, Kyoto University, Kitashirakawa-Oiwake-cho, Sakyo-ku, Kyoto 606-8502, Japan\\
$^{2}$National Astronomical Observatory of Japan, 2-21-1 Osawa, Mitaka, Tokyo, 181-8588, Japan\\
$^{3}$The Graduate University for Advanced Studies, 2-21-1 Osawa, Mitaka, Tokyo, 181-8588, Japan}

\begin{document}

\date{Accepted 2012 XX XX. Received 2012 XX XX; in original form 2011 November 30}

\pagerange{\pageref{firstpage}--\pageref{lastpage}} \pubyear{2002}

\maketitle

\label{firstpage}

\begin{abstract}
Recent observations of $z\ga7$ Ly$\alpha$ emitters (LAEs) have derived a variety of Ly$\alpha$ luminosity functions (LFs) with contradictory results, evolution or non-evolution from $z\la 6$, the epoch after reionization. This could be because most of $z\ga7$ LFs comprise photometric candidates and might include some contaminations. We conducted the Subaru Telescope Faint Object Camera And Spectrograph narrowband NB980 ($\lambda_c \sim 9800$\AA, FWHM $\sim 100$\AA) imaging and spectroscopy survey of $z=7$--7.1 LAEs to compare its "contamination-free" result with $z\ga7$ photometric Ly$\alpha$ LFs previously derived. We imaged the Subaru Deep Field and the sky around a cluster MS 1520.1+3002 and found one LAE candidate, but spectroscopy did not reveal Ly$\alpha$ though deep enough to detect it. We calculated the expected number of LAEs in our survey, using five $z=7$ and three $z=7.7$ Ly$\alpha$ LFs from recent surveys. Seven of them are consistent with null detection ($0.1^{+1.8}_{-0.1}$--$1.1^{+2.2}_{-1.0}$ LAEs) within errors including Poisson statistics and cosmic variance, but average values (0.7--1.1 LAEs) predicted from one $z=7$ and two $z=7.7$ LFs among the seven indicate nearly a single detection. The remaining one $z=7$ LF predicts $3.0^{+3.2}_{-2.0}$ LAEs. As to $z=7$, the discrepancy likely comes from different LAE selection criteria. For $z=7.7$, there are two possibilities; (1) If $z=7.7$ LAEs are somehow brighter in Ly$\alpha$ luminosity than lower redshift LAEs, $z=7.7$ LF is observed to be similar to or higher than lower redshift LFs even if attenuated by neutral hydrogen. (2) All/most of the $z=7.7$ candidates are not LAEs. This supports the decline of LF from $z\sim6$ to 7.7 and reionization at $z\sim6$--7.7.
\end{abstract}


\section{Introduction \label{Intro}}
Ly$\alpha$ emitters (LAEs) can be a probe of cosmic reionization, since their Ly$\alpha$ emission is absorbed or scattered by neutral hydrogen if the universe is not completely ionized, causing Ly$\alpha$ luminosity function (LF) to decline as the fraction of neutral hydrogen in intergalactic medium (IGM) increases \citep{RM01}. From $z=3$ to 5.7, the Ly$\alpha$ LF was observed not to evolve \citep[e.g.,][]{Ouchi08}. Also, many authors constructed statistically large and uniform samples of $z=5.7$ and 6.6 LAEs with some candidates spectroscopically confirmed \citep{Shimasaku06,Kashikawa06,Ouchi08,Ouchi10,Nakamura11}. They found that the Ly$\alpha$ LF significantly declines from $z=5.7$ to 6.6, suggesting that the universe could be partly neutral at $z=6.6$. The decline of the LF was further supported with large spectroscopic samples newly obtained by independent observations of \citet{Hu10} and \citet{Kashikawa11}. Meanwhile, \citet{Ota08,Ota10} found that Ly$\alpha$ LF also declines from $z=5.7$--6.6 to 7. These studies all imply that neutral fraction might increase with redshift at $z>6$. 

Moreover, \citet{Hu10} noticed that average Ly$\alpha$ equivalent width (EW) is slightly smaller at $z=6.6$ than 5.7. \citet{Kashikawa11} also found that Ly$\alpha$ EW distributions of $z \sim 3$--5.7 LAEs are similar, but EWs of $z=6.6$ LAEs are smaller. On the other hand, some Lyman break galaxies (LBGs) are known to show strong Ly$\alpha$ emission, while some do not. Fraction of Ly$\alpha$ emitting LBGs increases from $z\sim3$ to 6 \citep{Stark11} but suddenly drops from $z\sim6$ to 7 \citep{Ono12,Pentericci11, Schenker12}. The lower EW and Ly$\alpha$ LBG fraction at $z>6$ could be due to the rapid evolution of neutral fraction from $z\sim6$ to 7. This is consistent with the idea of the partly neutral Universe suggested by the decline of the Ly$\alpha$ LF.

However, some authors found a fair number of $z=7$ and 7.7 LAE candidates and claim that the Ly$\alpha$ LF does not evolve from $z=5.7$--6.6 to 7--7.7 \citep{Hibon10,Hibon11,Hibon12,Tilvi10,Krug12}. Their results contradict late reionization at $z<7.7$. Conversely, \citet{Clement11} also surveyed $z=7.7$ LAEs but did not detect any LAEs and support reionization at $z<7.7$. Interestingly, all the LFs implying no evolution at $z\ga6$ are based on photometric samples. If they suffer some contaminations, the conclusion would be different. The best way to reveal this is to identify all the candidates by spectroscopy. However, this requires very expensive campaigns that observe many candidates spread over different sky locations with sufficiently long integration to ensure detections or non-detections of Ly$\alpha$. Alternatively, simpler but indirect method is to conduct one imaging and spectroscopy survey and compare its {\it contamination free} result with the photometric LFs.

\begin{figure}
\includegraphics[width=62mm,angle=-90]{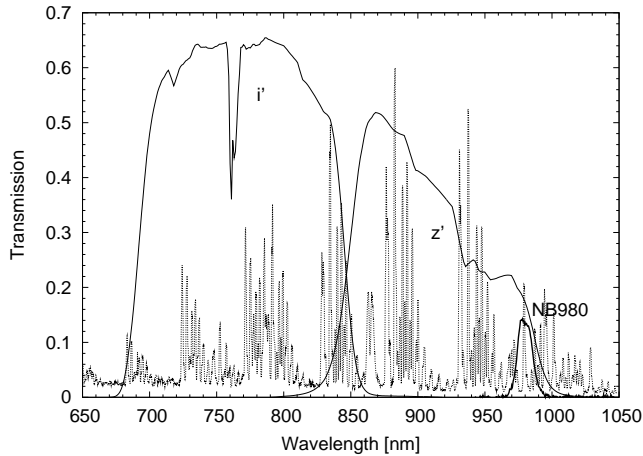}
%\vspace{130pt}
\caption{Transmission of the Subaru FOCAS NB980 and the Suprime-Cam $i'$ and $z'$ bands (solid curves). These curves include the CCD quantum efficiencies, the throughput of the telescope and instrument optics, and atmospheric transmission. The OH skylines are overplotted (dotted line).}
\label{FilterTransmission}
\end{figure}

Here, we use our imaging and spectroscopy survey of $z=7$--7.1 LAEs for this purpose. We did this survey in 2003, using the Subaru Telescope Faint Object Camera And Spectrograph \citep[FOCAS,][]{Kashikawa02} and a narrowband filter, NB980 ($\lambda_c \sim9800$\AA, ${\rm FWHM} \sim 100$\AA; see Figure \ref{FilterTransmission}) as a groundwork for our subsequent larger $z=7$ LAE surveys in 2005--2010 by \citet{Iye06} and \citet{Ota08,Ota10}. Since no detailed information about $z\sim7$ galaxies was available in 2003, we selected LAE candidates with tentative color criteria and conducted spectroscopy of them. In this paper, we re-analyze the NB980 data, refine the color criteria based on recent knowledge of $z\ga7$ galaxies, select $z=7$--7.1 LAE candidates and see if their spectra were taken in 2003 to derive a contamination-free result. 

In Section 2, we describe our imaging data. We perform selection of LAE candidates in Section 3. In Section 4, we explain the result of follow-up spectroscopy. In Section 5, we compare our NB980 survey with previously derived $z\ga7$ Ly$\alpha$ photometric LFs. We conclude in Section 6. Throughout we use an $(\Omega_m, \Omega_{\Lambda}, h) = (0.3, 0.7, 0.7)$ cosmology and $2''$ aperture AB magnitudes, unless otherwise specified.

%%% Figure 1
%\setcounter{figure}{12}

%%% Figure 2
%\setcounter{figure}{12}
\begin{figure}
\includegraphics[width=85mm,angle=0]{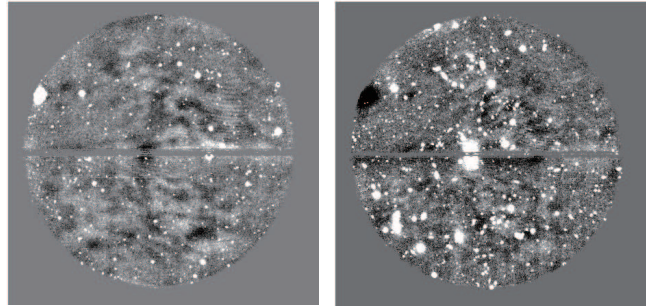}
%\vspace{130pt}
\caption{NB980-SDF (left) and NB980-MS1520 (right) images taken with two CCDs of the FOCAS. The north is up and the east to the left. The field of view is a circle of $6'$ diameter.}
\label{NB980images}
\end{figure}

%%% Figure 3
%\setcounter{figure}{12}
\begin{figure*}
\centering
\begin{minipage}{175mm}
\includegraphics[width=62mm,angle=-90]{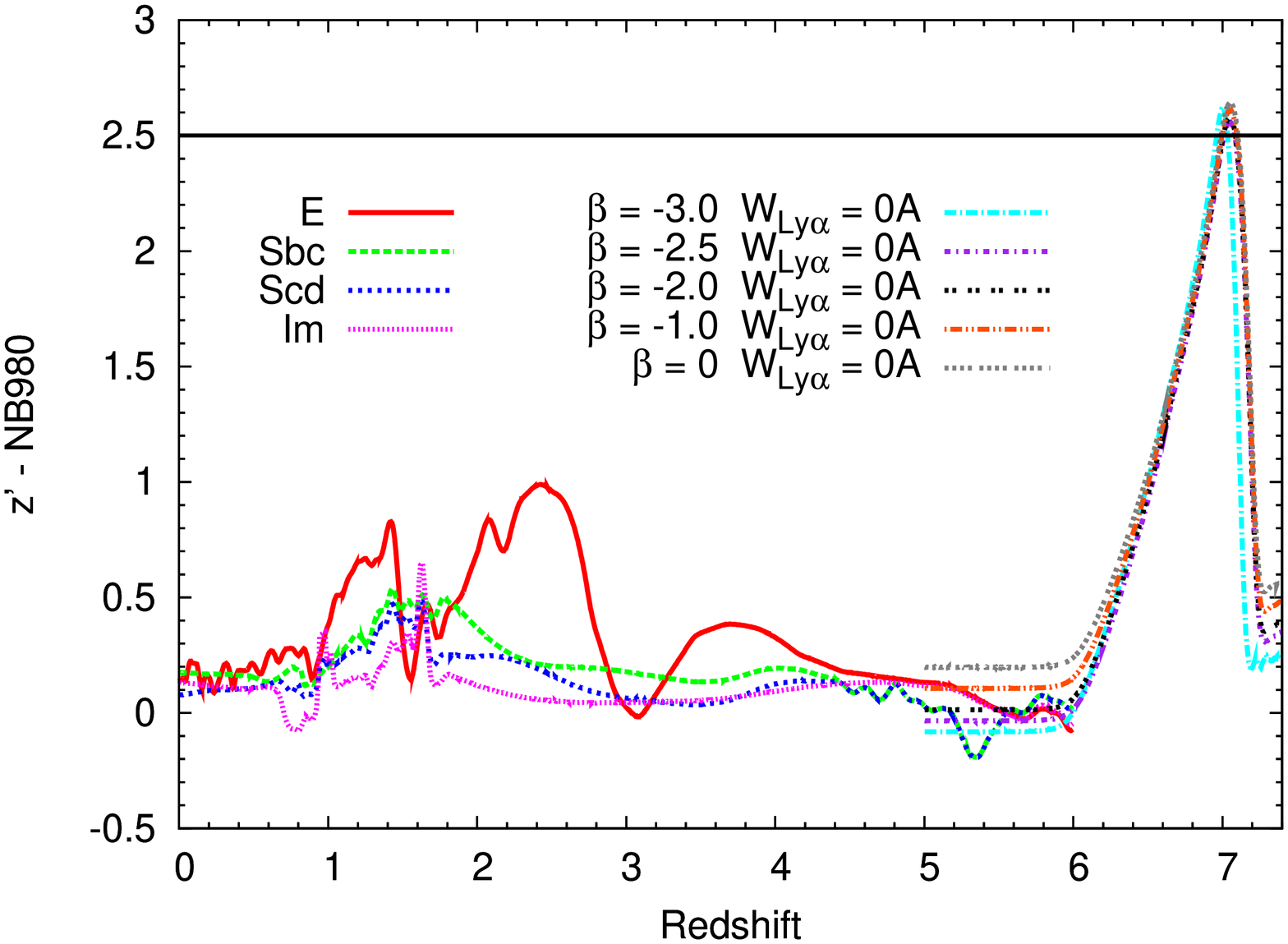}
\includegraphics[width=62mm,angle=-90]{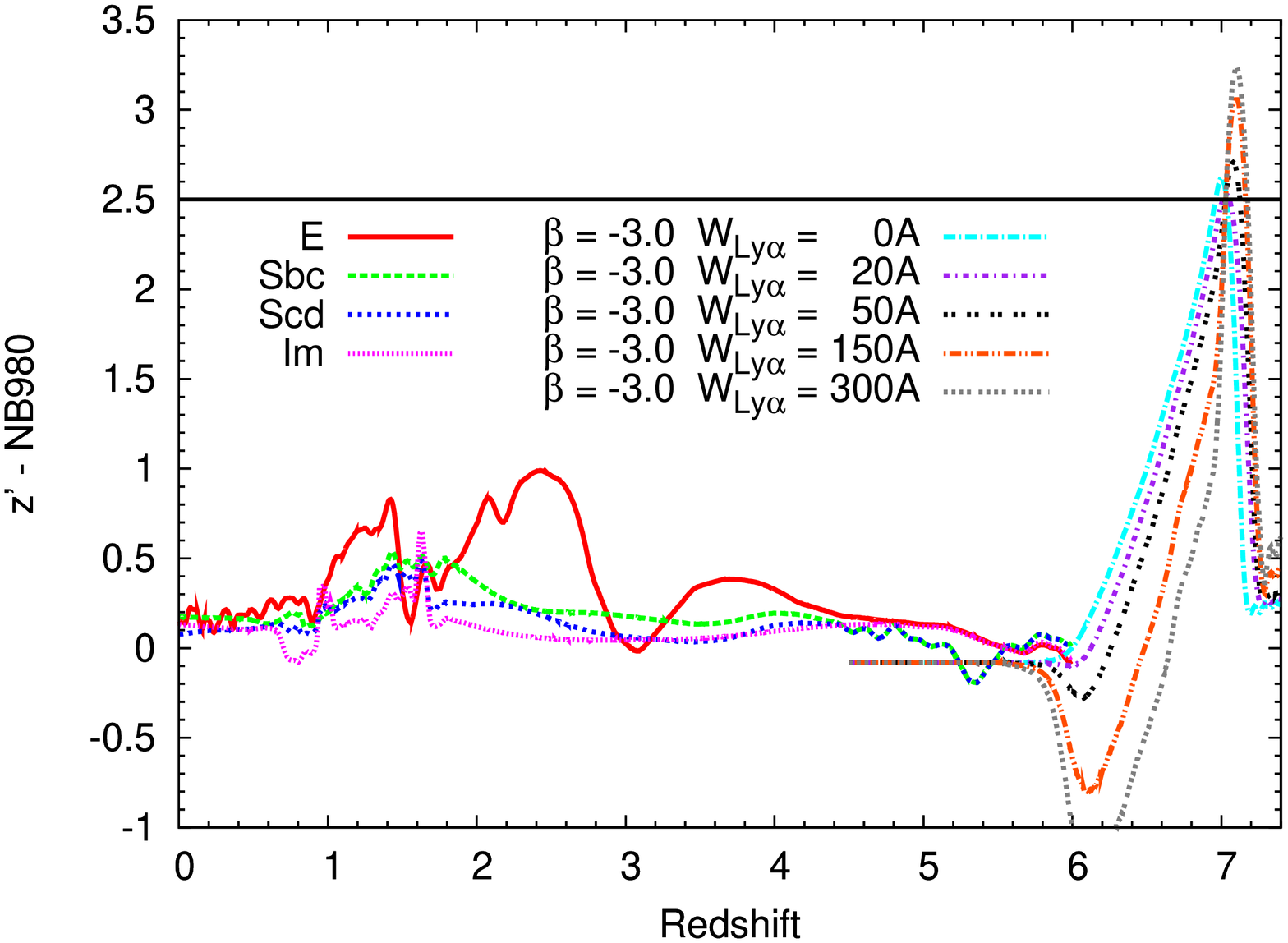}
\caption{Redshift evolution of $z' - {\rm NB980}$ colors of various galaxies. The colors of E (elliptical), Sbc, Scd and Im (irregular) galaxies were calculated using \citet{CWW80} template spectra. The colors of LBGs/LAEs were calculated using the spectrum $f_{\lambda} \propto \lambda^{\beta}$ for several UV continuum slopes $\beta$ and Ly$\alpha$ EWs $W_{{\rm Ly}\alpha}^{\rm rest}$. ({\it Left}) Colors of LBGs ($W_{{\rm Ly}\alpha}^{\rm rest}=0$). They show excess of $z' - {\rm NB980} \ga 2.5$ at $z\sim7$--7.1 (horizontal line), but the maximum excess at $z\sim7$--7.05 does not depend on $\beta$. ({\it Right}) Colors of an LBG and LAEs with $\beta=-3$ are plotted as a representative, as \citet{Ono10} suggest that $z\sim7$ LBGs and $z=5.7$ and 6.6 LAEs tend to have $\beta\simeq-3$. The LAEs clearly show the excess of $z' - {\rm NB980} > 2.5$ at $z\sim7$--7.1, which we adopted as the LAE selection criterion (3) in Section \ref{SelectionCriteria}. This criterion can detect both $z \sim 7$--7.1 LBGs ($W_{{\rm Ly}\alpha}^{\rm rest}< 20$\AA) and LAEs ($W_{{\rm Ly}\alpha}^{\rm rest}\ge 20$\AA).}
\label{CandidateColor1}
\end{minipage}
%\vspace{130pt}
\end{figure*}

\section{Imaging Observation and Data}
\subsection{Broadband and Narrowband Images}
The NB980 survey targeted the Subaru Deep Field \citep[SDF,][]{Kashikawa04} and the sky region around a galaxy cluster MS1520.1+3002 (hereafter, MS1520). For the SDF, broadband $BVRi'z'$ and narrowband NB816 ($\lambda_c=8150$\AA, ${\rm FWHM} = 120$\AA) and NB921 ($\lambda_c=9196$\AA, ${\rm FWHM} = 132$\AA) images were taken with the Subaru Telescope Suprime-Cam \citep{Miyazaki02} by the SDF project and NB973 ($\lambda_c=9755$\AA, ${\rm FWHM} = 200$\AA) image by \citet{Iye06}. All the SDF images were convolved to have a common point spread function (PSF) of $0\farcs98$. Limiting magnitudes at $3\sigma$ with $2''$ diameter apertures are ($B$, $V$, $R$, $i'$, $z'$, NB816, NB921, NB973) $=$ (28.45, 27.74, 27.80, 27.43, 26.62, 26.63, 26.54, 25.47). The $i'$ and $z'$ images of MS1520 were also taken by the Suprime-Cam and have PSFs of $0\farcs66$ and $0\farcs83$ and limiting magnitudes ($2''$ aperture, $3\sigma$) of 27.33 and 25.74. 

\subsection{NB980 Imaging and Data Reduction\label{NBreduction}}
We imaged the SDF and MS1520 with NB980 and FOCAS on 2003 May 7--8. The seeing was $0\farcs7$--$1\farcs2$. The 600 sec exposures were dithered with a simple pattern to minimize loss of the survey area. First, every time each exposure was taken, the pointing was shifted by $2''$ in the RA direction. After repeating this 8 (5) times for the SDF (MS1520), the pointing was returned to the original position in the RA direction but moved by $3''$ in the DEC direction. We repeated this procedure 3 times. The total integration times were 4 (SDF) and 2.7 (MS1520) hours. 

We reduced the NB980 data in the same manners as in \citet{Taka03}. The dithered exposures were registered and combined to produce final stacked images of the SDF and MS1520 (hereafter NB980-SDF and NB980-MS1520) as shown in Figure \ref{NB980images}. This reduced the fringing, but slight residual remained in the stacked images. The standard star Hz44 \citep{Oke90} was imaged during the observations to calibrate the photometric zeropoints. They were NB980 $=$ 30.27 mag ADU$^{-1}$ for both NB980-SDF and NB980-MS1520. The PSFs and $2''$ aperture $3\sigma$ limiting magnitudes of NB980-SDF and NB980-MS1520 were $1\farcs25$ and $1\farcs27$, and NB980 $=24.67$ and 24.44, respectively. Assuming the minimum detectable rest frame Ly$\alpha$ EW of $W_{{\rm Ly}\alpha}^{\rm rest}=20$\AA~(see Section \ref{SelectionCriteria}), these magnitudes correspond to Ly$\alpha$ flux limits of $F_{{\rm Ly}\alpha} \sim 1.2$ and $1.4 \times 10^{-17}$ erg s$^{-1}$ cm$^{-2}$ or Ly$\alpha$ luminosity limits of $L_{{\rm Ly}\alpha} \sim 6.7$ and $8.3 \times 10^{42}$ erg s$^{-1}$, respectively. The effective area imaged with the FOCAS and NB980 was $\sim 56$ arcmin$^2$ (SDF plus MS1520). The comoving distance along the line of sight corresponding to the redshift $7.0 \leq z \leq 7.1$ probed by the NB980 was $\sim 29.1 h^{-1}$Mpc. Hence, we surveyed a comoving volume of $\sim10^4$ Mpc$^3$.  

%%% Figure 4
%\setcounter{figure}{12}
\begin{figure}
\includegraphics[width=85mm,angle=0]{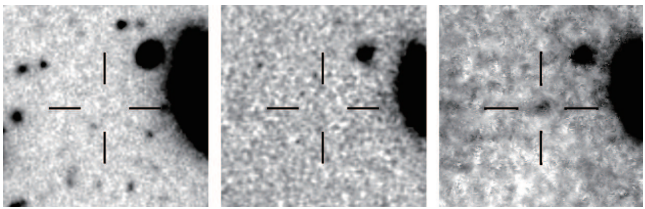}\\
\caption{The $20\arcsec \times 20\arcsec$ $i'$, $z'$ and NB980 images (from left to right) of MS1520-A. The north is up, and the east is to the left.}
\label{VIzNBJH}
\end{figure}

\section{LAE Candidate Selection}
\subsection{Photometry}
We first matched astrometry and pixel scales of all the Suprime-Cam images to those of NB980-SDF and NB980-MS1520. Also, we convolved PSFs of the $z'$-band images to those of NB980-SDF and NB980-MS1520 to calculate $z' - {\rm NB980}$ color by measuring $z'$ and NB980 magnitudes with the same aperture to select LAE candidates in Section \ref{SelectionCriteria}. Then, we performed source detection and photometry with the SExtractor \citep{BA96}. The pixel scale of the NB980 images is $0\farcs1038$ pixel$^{-1}$. We regarded an area larger than 5 contiguous pixels with a flux greater than $2\sigma$ as an object. We detected objects in NB980-SDF and NB980-MS1520 and did photometry in other wavebands, using the double-imaging mode. The $2''$ diameter aperture magnitudes were measured with the {\tt MAG\_APER} parameter and total magnitudes the {\tt MAG\_AUTO}. Finally, combining the photometry in all the wavebands, we constructed the NB980-detected object catalogs for the SDF and MS1520.

\subsection{LAE Criteria and Candidate Selection\label{SelectionCriteria}}
Figure \ref{FilterTransmission} shows that the NB980 band is located at the red side of the $z'$ band. If the Lyman break of an LBG is redshifted into NB980, it results in a significant excess of $z'-{\rm NB980}$. If the spectrum has Ly$\alpha$ emission, the excess is more significant. We used this characteristic to isolate $z\sim7$ LBGs/LAEs. We examined the expected $z' - {\rm NB980}$ colors of $z\sim7$ LBGs/LAEs and derived candidate selection criteria. We first created model spectra of LBGs having the power law continua $f_{\lambda} \propto \lambda^{\beta}$ with several different slopes $\beta=-3, -2.5, -2, -1.5, -1, 0$ and then model spectra of LAEs by adding Ly$\alpha$ emission with rest frame EWs of $W_{\rm Ly\alpha}^{\rm rest}=20, 50, 100, 150$ and 300\AA. We did not assume any specific line profile or velocity dispersion. Instead, we simply added the total line flux value to the spectra at 1216\AA. Then, we redshifted the spectra to $z=0$--8 and applied Ly$\alpha$ absorption by IGM, using the prescription of \citet{Madau95}. 

Colors of LBGs/LAEs were calculated using the model spectra and transmission curves of $z'$ and NB980 and plotted as a function of redshift in Figure \ref{CandidateColor1}. For comparison, we also calculated the colors of E (elliptical), Sbc, Scd and Im (irregular) galaxies using the \citet{CWW80} template spectra. As clearly seen, LBGs ($W_{\rm Ly\alpha}^{\rm rest}=0$) and LAEs ($W_{\rm Ly\alpha}^{\rm rest} \ge 20$\AA) are expected to produce significant excess in NB980 against $z'$ at $z\sim7$--7.1. Also, $z \sim 1$--3 ellipticals show modest excess due to the 4000\AA~Balmer break. We adopted $z' - {\rm NB980} > 2.5$ as a criterion that selects $z\sim7$--7.1 LBGs/LAEs avoiding ellipticals. We used the following criteria to select LBG/LAE candidates. 
\begin{eqnarray}
&&{\rm NB980} \le 3\sigma~{\rm limiting~magnitude}\\ 
&&{\rm Bands~blueward~of~Ly}\alpha < 2\sigma,~{\rm and~no~object~is~seen}\nonumber\\
&&{\rm visually}\\
&&z' - {\rm NB980} > 2.5\\
&&{\rm NB973} < 2\sigma~{\rm limiting~magnitude},~{\rm and}\nonumber\\
&&{\rm the~object~is~seen~visually~(only~for~the~SDF)}
\end{eqnarray}
The (1) is the object detection limit. The (2) means nondetections in $B$, $V$, $R$, $i'$, NB816 and NB921 for SDF and in $i'$ for MS1520. No object should be seen by eyes in any of these bands since the flux blueward of Ly$\alpha$ should be absorbed by IGM. This eliminates interlopers such as L/M/T type stars and low redshift galaxies such as H$\beta$, [OIII], [OII], H$\alpha$ or [SII] line emitters. The (3) is the NB980 excess as explained above. In calculating the color, if the $z'$ magnitude was fainter than $1\sigma$, it was replaced by the $1\sigma$ value. The (4) is ${\rm NB973} < 25.9$ meaning a detection in NB973 whose bandpass targets $z=6.9$--7.1 Ly$\alpha$. We have the NB973 image of the SDF. Because the NB973 image is deeper than the NB980 image, any object detected in the NB980 image should be detected and visually seen in the NB973 image. 

We applied the criteria (1)--(4) to the NB980-detected object catalogs. Then, we visually inspected the NB980 images of selected objects to remove obviously spurious ones such as tails of saturated pixels from bright stars, halos of bright stars, noises  of anomalously high fluxes at the edges of CCDs and field of view, and residual of fringing. As a result, one object in MS1520 remained (hereafter MS1520-A). Table 1 and Figure 4 show its photometry and images. 

% Table 1
\begin{table}
%\centering
\begin{minipage}{85mm}
\caption{Photometry of MS1520-A.}
\begin{tabular}{@{}cccccc@{}}
\hline 
RA(J2000) & DEC(J2000) & $i'$ & $z'$ & NB980 & NB980\\
\hline 
15:22:11.8 & +29:50:40.3 & $>$28.52 & $>$26.94 & 23.87 & 23.42\\    
\hline     
\end{tabular}\\
NOTE: The columns 3--5 (6) are $2''$ aperture (total) magnitudes. The limits are $1\sigma$.  
\end{minipage}
\end{table}

\section{Follow-up Spectroscopy}
In our previous NB980 survey in 2003, we had selected more candidates with more inclusive criteria $z' - {\rm NB980} > 1.3$ and $2\sigma$ limiting magnitudes. Then, we had conducted spectroscopy on 2003 June 22--23 with multi-object slits and FOCAS. We confirmed that MS1520-A selected in Section \ref{SelectionCriteria} had been also observed in 2003. At that time, we had used VPH950 grism (grating of 1095 lines mm$^{-1}$ and resolution of $\sim 2500$) with O58 order-cut filter (coverage of 580--1000 nm) and $0\farcs8$ slits. Integration time was 4 hours comprising six 2400 seconds exposures dithered along the slit by $\pm1''$. The spectra of the standard star Feige 34 \citep{Oke90} had been also obtained and used for flux calibration. The data reduction was performed in the same standard manners as in \cite{Iye06}. We inspected the sky-subtracted stacked spectrum of MS1520-A and could identify neither Ly$\alpha$ emission, a UV continuum nor any other spectral features. 

\subsection{Possibility of the Candidate Being an LAE\label{LAEPossibility}} 
To see if we had enough depth to detect Ly$\alpha$, we compared the sky background RMS of the stacked spectrum with Ly$\alpha$ flux estimated from NB980 magnitude. If we assume $z=7$ and a rest frame Ly$\alpha$ EW as low as $W_{{\rm Ly}\alpha}^{\rm rest}=20$\AA~for the most severe case, $\sim 76$\% of the NB980 flux comes from Ly$\alpha$ line, and total NB980 magnitude of MS1520-A, 23.42 converts to Ly$\alpha$ flux, $F^{\rm NB}_{{\rm Ly}\alpha}=3.7 \times 10^{-17}$ erg s$^{-1}$ cm$^{-2}$. \citet{Ota10b} estimated that slit loss of the flux in the FOCAS spectroscopy of a $z=6.96$ LAE IOK-1 \citep{Iye06} is $\sim35$\% with a $0\farcs8$ slit under a seeing of $1''$. We also used $0\farcs8$ slits for our spectroscopy, and the seeing was also $\sim 1''$. If we apply this slit loss, the flux is $F^{\rm NB}_{{\rm Ly}\alpha}=2.4\times 10^{-17}$ erg s$^{-1}$ cm$^{-2}$. Meanwhile, we used binning of 4 pixels (equivalent to $0\farcs4$, smaller than the seeing of $\sim1''$) in the spatial direction to extract one dimensional spectrum of MS1520-A. Calculating the variance in unbinned pixels in the dispersion direction at 9750--9850\AA~(NB980 passband) in this spectrum, we estimated the sky RMS to be $8.8 \times 10^{-19}$ erg s$^{-1}$ cm$^{-2}$ \AA$^{-1}$. The FWHM of Ly$\alpha$ line, for example, of a $z=6.6$ LAE varies from 5.5 to 14.6\AA~\citep{Kashikawa06,Taniguchi05}. If we assume the FWHMs of $z=7$ LAEs are similar, the Ly$\alpha$ flux is $F^{\rm spec}_{{\rm Ly}\alpha}=(0.48$--$1.3)\times 10^{-17}$ erg s$^{-1}$ cm$^{-2}$. This is 2--5 times fainter than $F^{\rm NB}_{\rm Ly\alpha}$, deep enough to detect Ly$\alpha$. Thus, MS1520-A is unlikely an LAE. 

\subsection{Possibility of the Candidate Being an LBG} 
Another possible origin of MS1520-A is a $z\sim7$ LBG. However, we could neither see any faint continuum by eye on the 2 dimensional spectrum nor in the 1 dimensional one after the 4 pixel binning in the spatial direction. To see if we had enough depth to detect the continuum, we compared the sky RMS of the stacked spectrum with the UV continuum flux density estimated from NB980 magnitude. If we assume $z=7.05$ Lyman break (i.e., all the NB980 flux are from the UV continuum at the red half of NB980 and zero flux otherwise), total NB980 magnitude of MS1520-A, 23.42 and 35\% slit loss converts to a UV continuum flux density, $f^{\rm NB}_{\lambda, {\rm UV}}=6.4\times 10^{-19}$ erg s$^{-1}$ cm$^{-2}$ \AA$^{-1}$. The sky RMS of the 1 dimensional spectrum, $8.8 \times 10^{-19}$ erg s$^{-1}$ cm$^{-2}$ \AA$^{-1}$, is 1.4 times shallower than $f^{\rm NB}_{\lambda, {\rm UV}}$, not enough to detect the continuum. Our spectroscopy does not reveal if MS1520-A is a $z\sim7$ LBG. 

Another way is to see if our NB980 imaging was deep enough to detect $z \sim7$ LBGs. \citet{Ouchi09} recently detected 22 $z\sim7$ LBG candidates in the SDF and the GOODS-N fields to the UV luminosity $M_{\rm UV}=-21$. If we assume that all the NB980 flux is from the UV continuum and that $z=7$, our depths, ${\rm NB980}=24.67$ (SDF) and 24.44 (MS1520), convert to $M_{\rm UV} \sim -22.3$ and $-22.5$. To these limits, \citet{Ouchi09} did not detect any LBGs. Hence, MS1520-A is unlikely a $z\sim7$ LBG. 

\subsection{Other Possibilities} 
The other possible origin of MS1520-A is a late-type star, a variable/transient object or a noise. In case of a late-type star, the spectrum could show no signal if the continuum is fainter than our spectroscopy limit. Because we have only one image blueward of Ly$\alpha$ ($i'$-band), the null detection criterion only on this band might not be strict enough to remove a faint late-type star. In case of a variable/transient object, it could have been fainter than our detection limit at the time of $i'$-band imaging and the spectroscopy while it might have been bright at the time of NB980 imaging. 

Meanwhile, MS1520-A is unlikely a low-$z$ line emitter. If it was, its line flux would be as bright as the $F^{\rm NB}_{{\rm Ly}\alpha}$ estimated in Section \ref{LAEPossibility} and detectable by the spectroscopy. 

The purpose of this study is to obtain a contamination free result for $z\sim7$ LAEs, including null detection. We can safely conclude that we detect no $z\sim7$ LAE in our survey volume and to our detection limit. We used this information to assess $z\ga7$ photometric Ly$\alpha$ LFs in the literatures.     

% Table 2
\begin{table*}
\centering
\begin{minipage}{177mm}
%\begin{minipage}{84mm}
\begin{center}
\caption{Expected detection number of LAEs in the NB980 survey estimated from $z=7$ and 7.7 Ly$\alpha$ LFs.}
\label{ExpLAEnum}
\begin{tabular}{@{}lcccccc@{}}
\hline
Authors	& Volume$^a$ & Ly$\alpha$ flux limit$^a$ & \#LAE candidates$^d$ & \multicolumn{3}{c}{\underline{Expected \#LAEs in NB980 survey$^e$}}\\		
		& (10$^4$ Mpc$^3$) & (10$^{-17}$ erg s$^{-1}$ cm$^{-2}$) & authors detected & SDF & MS1520 & SDF+MS1520\\
\hline
\multicolumn{7}{c}{$z=7$ LAEs}\\		
\hline	
This study              & 1.0 & 1.2--1.4 & 0 & ---     & ---      & ---\\				
\citet{Ota08}           & 32  & 1.5$^b$  & 1 & $0.3^{+1.9}_{-0.3}$ & $0.2^{+1.8}_{-0.2}$   & $0.3^{+1.8}_{-0.3}$\\ 
\citet{Ota10}           & 30  & 0.97$^b$ & 3 & $0.1^{+1.8}_{-0.1}$ & $0.07^{+1.8}_{-0.07}$ & $0.1^{+1.8}_{-0.1}$\\ 
\citet{Hibon11}         & 7.2 & 1.8$^c$  & 6 & $2.0^{+2.8}_{-1.6}$ & $1.5^{+1.9}_{-1.5}$   & $3.0^{+3.2}_{-2.0}$\\ 
\citet{Hibon12} D33$^f$ & 40  & 2.7$^b$  & 7 & $0.5^{+1.9}_{-0.5}$ & $0.2^{+1.8}_{-0.2}$   & $0.5^{+1.9}_{-0.5}$\\ 
\citet{Hibon12} D41$^f$ & 43  & 1.9$^b$  & 7 & $1.1^{+2.3}_{-1.1}$ & $0.5^{+1.9}_{-0.5}$   & $1.1^{+2.2}_{-1.0}$\\ 
\citet{Kobayashi07}     & --- & ---    & --- & $0.4^{+1.9}_{-0.4}$ & $0.3^{+1.8}_{-0.3}$   & $0.6^{+2.7}_{-0.5}$\\ 
\hline
\multicolumn{7}{c}{$z=7.7$ LAEs}\\	   
\hline						
\citet{Hibon10} & 6.3 & 0.83 & 7 & $0.3^{+1.9}_{-0.3}$ & $0.2^{+1.8}_{-0.2}$ & $0.3^{+1.8}_{-0.3}$\\ 
\citet{Tilvi10} & 1.4 & 0.60 & 4 & $0.7^{+2.6}_{-0.6}$ & $0.4^{+1.9}_{-0.4}$ & $0.8^{+2.5}_{-0.7}$\\
\citet{Krug12}  & 2.8 & 0.80 & 4 & $0.5^{+1.9}_{-0.5}$ & $0.4^{+1.9}_{-0.4}$ & $0.7^{+2.6}_{-0.6}$\\ 
\hline
\end{tabular}
\end{center} 
$^a$The survey volume and limit of the papers in the column 1. 
$^b$Ly$\alpha$ fluxes converted from their limiting NB973 magnitudes ($5\sigma$, $2''$ aperture), assuming a rest frame Ly$\alpha$ EW of 20\AA~and $z=7$.  
$^c$The Ly$\alpha$ flux directly converted from the NB9680 magnitude of their faintest LAE candidate.
$^d$The number of LAE candidates the authors detected in their survey volumes and to their survey limits.
$^e$The expected detection number of LAEs in our NB980 survey estimated using the Ly$\alpha$ LFs from the papers in the column 1. 
For \citet{Ota08}, we integrated the inferred $z=7$ Ly$\alpha$ LF in Figure 10 of their paper. \citet{Ota10} Ly$\alpha$ LF is based on $L_{{\rm Ly}\alpha}$ directly converted from Ly$\alpha$ fluxes of their candidates. \citet{Ota10b} estimated that $\sim77$\% of NB973 total flux of a $z=6.96$ LAE (IOK-1) is from Ly$\alpha$. Hence, we interpolated \citet{Ota10} Ly$\alpha$ LF attenuated by $0.77 \times L_{{\rm Ly}\alpha}$ to our NB980 survey limits to obtain the expected number of LAEs. For \citet{Hibon10,Hibon11,Hibon12}, we integrated the \citet{Schechter76} LFs best-fitted to their observed data by them. For \citet{Kobayashi07}, we integrated the $T_{{\rm Ly}\alpha}^{\rm IGM}=1$ Ly$\alpha$ LF predicted by their model. For \citet{Tilvi10} and \citet{Krug12}, we interpolated their Ly$\alpha$ LFs to our NB980 survey limits. The errors include Poisson errors for small number statistics \citep{Gehrels86} and cosmic variance $\sigma_v$. We combined these errors quadratically and if the lower limit was a negative value, we adjusted it to zero. We calculated $\sigma_v$'s using the bias $b=3.4$ obtained for LAEs by \citet{Ouchi05} and the dark matter halo variances $\sigma_{\rm DM}$'s at $z=6$ predicted by the analytic model of \citet{Somerville04} and our survey volumes. The $\sigma_v$'s are $\sim 48$\% ($\sim 54$\%) for the survey volumes equivalent to two (one) FOCAS field of views. 
$^f$The Schechter parameters $\Phi^*$ and $L^*$ for LFs derived from the D33 and the D41 fields in Table 4 in \citet{Hibon12} paper was found to be mistakenly listed. They are switched with each other (P. Hibon 2012, private communication). We used their Schechter LFs with correct parameters to estimate the expected LAE numbers. 
\end{minipage}
\end{table*} 

\section{Discussion}
We compared our null detection with the expected detection number of LAEs estimated from a variety of $z=7$ and 7.7 photometric Ly$\alpha$ LFs to check their consistency with the current contamination free result. 

\subsection{Comparison with $z=7$ Ly$\alpha$ LFs\label{z7LF}}
In Table \ref{ExpLAEnum}, we calculated and listed expected detection number of LAEs in our NB980 survey volume by integrating or interpolating Ly$\alpha$ LFs from four recent $z=7$ LAE surveys \citep{Ota08,Ota10,Hibon11,Hibon12} to our survey limits. The errors include Poisson error for small number statistics and cosmic variance (see the footnote of Table 2 for details). For comparison, we also calculated the expected number in the case of no neutral hydrogen attenuation of Ly$\alpha$ emission (i.e., Ly$\alpha$ transmission to neutral hydrogen $T_{{\rm Ly}\alpha}^{\rm IGM}=1$) by integrating such a $z=7$ Ly$\alpha$ LF predicted by a recent LAE evolution model of \citet{Kobayashi07} to our survey limits. The expected numbers estimated using LFs from \citet{Ota08,Ota10} and one of the \cite{Hibon12} survey fields (D33) are consistent with the null detection. Conversely, the number estimated using \citet{Hibon11} LF indicates the detection of $3.0^{+3.2}_{-2.0}$ LAEs. Also, though consistent with null detection within the error, the number, $1.1^{+2.2}_{-1.0}$, estimated from another LF (D41) of \citet{Hibon12} is larger than those estimated from other $z=7$ LFs. This implies that LAE candidates detected by \citet{Hibon11,Hibon12} might include some degree of contaminations. We compared \citet{Ota08,Ota10} and \citet{Hibon11,Hibon12} surveys and investigated what causes the discrepancy.    

\citet{Ota08,Ota10} obtained the narrowband NB973 images of the SDF and the SXDS \citep{Furusawa08} fields with one pointing of the Subaru Suprime-Cam  each, reaching limiting magnitudes ${\rm NB973}= 24.9$ and 25.4 ($5\sigma$, $2''$ aperture). Meanwhile, \citet{Hibon12} imaged two fields called D33 and D41 with the same NB973 filter and one Suprime-Cam pointing each and reached limiting magnitudes ${\rm NB973} = 24.3$ and 24.7 ($5\sigma$, $2''$ aperture). Though \citet{Hibon12} images are shallower than \citet{Ota08,Ota10}, they detected 7 and 7 LAE candidates in the D33 and the D41, while \citet{Ota08,Ota10} detected 1 and 3 in the SDF and the SXDS (see Table \ref{ExpLAEnum}). Moreover, \citet{Hibon11} imaged the COSMOS field with their narrowband NB9680 ($\lambda_c=9680$\AA, FWHM$\sim90$\AA) to detect 6 LAE candidates, though their limit and volume are shallower and smaller than those of \citet{Ota08,Ota10}.

Meanwhile, \citet{Kobayashi07} $z=7$ LF with $T_{{\rm Ly}\alpha}^{\rm IGM}=1$ predicts the detection of $0.6^{+2.7}_{-0.5}$ LAEs even if the neutral fraction at $z=7$ is 0\%. Because \citet{Hibon11,Hibon12} LFs predict $1.1^{+2.2}_{-1.0}$--$3.0^{+3.2}_{-2.0}$ detections whether or not neutral fractions is 0\%, the difference in the expected LAE number between \citet{Hibon11,Hibon12} and \citet{Ota08,Ota10} cannot be explained by field-to-field variation of the degree of Ly$\alpha$ attenuation by neutral hydrogen. In addition, since \citet{Hibon12} and \citet{Ota08,Ota10} surveyed similar and very large volumes, cosmic variance and Poisson error are not the cause of difference, either. Moreover, because \citet{Ota10} probed to much deeper limit than \citet{Hibon11,Hibon12}, if {\it difference} in dust extinction of all the $z=7$ LAEs in different sky fields is $E(B-V) < 0.06$ (equivalent to difference in survey depth), dust extinction is also unlikely the reason. For example, \citet{Ono12} constrained dust extinction of a spectroscopically confirmed $z=7.213$ LAE to be $E(B-V) \sim 0.05$. Though it is one example, dust extinction of individual $z\sim7$ LAEs seem to be modest. Thus {\it difference} in dust extinction among $z\sim7$ LAEs could be even smaller. One remaining factor that causes the discrepancy between \citet{Ota08,Ota10} and \citet{Hibon11,Hibon12} is different LAE selection criteria. 
 																
\subsubsection{Effect of Selection Criteria on Detection Number} 
\citet{Hibon11,Hibon12} did not consider the redshift evolution of narrowband excess of galaxies, while \citet{Ota08,Ota10} did. The NB9680 and NB973 they used have FWHMs of 90\AA~and 200\AA, and located at the red edge of $z'$-band. This makes $z<7$ LAEs/LBGs also selected as candidates. In the case of NB973, $z\ga6.4$ Ly$\alpha$ emission with EWs of even $W_{{\rm Ly}\alpha}^{\rm rest} \la 20$\AA~can produce significant excess of $z'-{\rm NB973} \ga 0.7$, according to Figure 3 of \citet{Ota08} that plots redshift evolution of $z' - {\rm NB973}$ colors of model LAEs/LBGs generated with \citet{BC03} population synthesis models. The same Figure also shows that a $z=7$ LAE/LBG is expected to have $z'-{\rm NB973} \ga 1.9$. We also calculated $z'-{\rm NB973}$ color versus redshift similar to Figure 3 in the present paper, using spectra $f_{\lambda} \propto \lambda ^{\beta}$ with $\beta=-3$ to 0 and $W_{{\rm Ly}\alpha}^{\rm rest}=0$ to 300\AA~and found that a $z=7$ LAE/LBG is expected to have $z'-{\rm NB973} \ga2.2$. For example, a spectroscopically confirmed $z=6.96$ LAE \citep[IOK-1,][]{Iye06} has a color of $z'-{\rm NB973} > 2.44$. \citet{Ota08,Ota10} adopted slightly less strict criterion $z'- {\rm NB973} > 1$--1.72, considering the effects of photometric errors and possible diversity of $z=7$ LAEs on the color. This criterion in principle selects $z\ga6.5$ LAEs/LBGs. Hence, \citet{Ota08,Ota10} imposed an additional criterion, a null detection in the narrowband NB921 ($\lambda_c=9196$\AA, ${\rm FWHM}=132$\AA) to avoid $z=6.5$--6.6 LAEs/LBGs. Hence, their criteria select $z\ga6.7$ LAEs/LBGs. 

Meanwhile, \citet{Hibon12} used even more inclusive criterion $z'-{\rm NB973} > 0.65$, which selects $z\ga 6.4$ LAEs/LBGs, and they did not have NB921 images to avoid $z=6.5$--6.6 LAEs. Their sample include 4 and 6 candidates with $z'-{\rm NB973} \sim 1$--1.5 in the D33 and the D41 fields. Some of them might be $z=6.5$--6.6 LAEs, which \citet{Ota08,Ota10} did not select. In fact, \citet{Ota08} reported that if they did not impose null detection in NB921, they selected a spectroscopically confirmed $z=6.6$ LAE IOK-3 with $z'-{\rm NB973} \sim 1.41$ as a candidate. Besides, \citet{Hibon11} adopted ${\rm NB9680} -z' < -0.75$ without examining redshift evolution of this color of LAEs. For the same reason as the case of NB973, \citet{Hibon11} sample might potentially include some lower redshift LAEs/LBGs. If \citet{Hibon11,Hibon12} candidates really include some such contaminations, their real $z=7$ Ly$\alpha$ LFs after corrected for the contaminations could show evolution from $z\sim6$ and predict more consistent number of LAEs in our NB980 survey.

\subsection{Comparison with $z=7.7$ Ly$\alpha$ LFs\label{z7p7LF}}
In Table \ref{ExpLAEnum}, we also listed the expected number of LAEs in our survey estimated using 3 $z=7.7$ Ly$\alpha$ LFs \citep{Hibon10,Tilvi10,Krug12} under the assumption that LF does not evolve from $z=7$ to 7.7. All the expected numbers are consistent with null detection within errors. Nonetheless, the average value 0.7--0.8 of the expected number ($0.7^{+2.6}_{-0.6}$--$0.8^{+2.5}_{-0.7}$ LAEs) estimated from the LFs of \citet{Tilvi10} and \cite{Krug12} is close to a single detection that means evolution of LF from $z=7$ to 7.7. Actually, these authors and even \citet{Hibon10} claim that if their candidates are all real $z=7.7$ LAEs, their LFs imply a trend opposite to the decline of Ly$\alpha$ LF from $z=5.7$ to 6.6 and 7 found by \citet{Kashikawa11}, \citet{Ouchi10} and \citet{Ota08,Ota10}. \citet{Tilvi10} and \cite{Krug12} also performed Monte Carlo simulations to estimate the expected number of LAEs in their surveys and concluded that even if 1 or 2 candidates in their samples are real $z=7.7$ LAEs, it implies no evolution of Ly$\alpha$ LF from $z=6.5$ to 7.7. There are two possible scenarios that explain the situation.

One is that $z=7.7$ LAEs might be somehow brighter than $z=6.5$ LAEs in Ly$\alpha$ luminosity. In this case, $z=7.7$ LF is observed to be similar to or higher than $z=6.5$ LF even if attenuated by neutral hydrogen. There are two factors potentially supporting this. First, stellar population studies of $z\sim6$ massive ($\sim 10^{10} M_{\odot}$) LBGs suggest that they could have had substantially high star formation rates in the past at $z\sim7$--8 to assemble their high masses and could be very bright at $z\ga7$ \citep{Yan06,Eyles07}. Also, narrowband selected LAEs tend to be the lower mass and younger age extension of LBG population and might be at the earlier stage of the evolution of LBGs \citep{Lai08,Pirzkal07}. If $z=7.7$ LAEs are the progenitors of $z\sim6$ massive LBGs, they might be intrinsically very bright. Second, \citet{Hayes11} estimated Ly$\alpha$ escape fractions $f_{\rm esc}^{{\rm Ly}\alpha}$ at $z\sim0$--7.7 using observed data in literatures and found that $f_{\rm esc}^{{\rm Ly}\alpha}$ increases with redshift when they fitted a power law to $f_{\rm esc}^{{\rm Ly}\alpha}$'s at $z\sim0$--6. If $f_{\rm esc}^{{\rm Ly}\alpha}$ is higher in LAEs at $z=7.7$ than 6.5, $z=7.7$ LAEs are observed to be brighter.

Another scenario is that all the LAE candidates of \citet{Tilvi10} and \cite{Krug12} and all/some of \citet{Hibon10} LAE candidates are not $z=7.7$ LAEs. This agrees with the decline of Ly$\alpha$ LF from $z=5.7$ to 6.6--7 found by \citet{Kashikawa11}, \citet{Ouchi10} and \citet{Ota08,Ota10}. This scenario is supported by another $z=7.7$ LAE survey by \citet{Clement11} who did not detect any candidates. As mentioned earlier, the drop in fraction of Ly$\alpha$ emitting LBGs at $z>6$ \citep{Ono12,Pentericci11,Schenker12} could be also a support. 

\section{Conclusion}
We conducted the Subaru FOCAS NB980 imaging and spectroscopy survey of $z=7$--7.1 LAEs in the SDF and MS1520 and detected no LAEs to a $3\sigma$ Ly$\alpha$ flux limit of $\sim 1.4\times10^{-17}$ erg s$^{-1}$ cm$^{-2}$ in a comoving volume of $\sim10^4$ Mpc$^3$.  

We estimated the expected number of LAEs in our survey from 5 $z=7$ and 3 $z=7.7$ photometric Ly$\alpha$ LFs. Seven agree with null detection within errors but average LAE numbers predicted by one $z=7$ and two $z=7.7$ LFs among the seven indicate nearly a single detection. The remaining one $z=7$ LF predicts a detection of $3.0^{+3.2}_{-2.0}$ LAEs. For $z=7$, the discrepancy likely comes from different LAE selection criteria. For $z=7.7$, there are 2 possibilities. (1) If $z=7.7$ LAEs are brighter than lower-$z$ LAEs, $z=7.7$ LF is similar to or higher than lower-$z$ LFs even if attenuated by neutral hydrogen. (2) If all the $z=7.7$ candidates are not real, Ly$\alpha$ LF declines from $z\sim6$ to 7.7.

Evaluation of $z\ga7$ Ly$\alpha$ LFs in this study relies on an indirect method. The direct and best way is secure spectroscopy of all the candidates. Future powerful telescopes such as JWST or TMT will facilitate this and reveal the real nature of Ly$\alpha$ LFs at the epoch of reionization.

\section*{Acknowledgments}
This work is based on data collected at the Subaru Telescope, which is operated by the National Astronomical Observatory of Japan. This work was supported by the Grant-in-Aid for the Global COE Program "The Next Generation of Physics, Spun from Universality and Emergence" from the Ministry of Education, Culture, Sports, Science and Technology (MEXT) of Japan. We thank our referee for useful comments that improved this paper. We thank Pascale Hibon, Nobunari Kashikawa and Takatoshi Shibuya for the information about \citet{Hibon12} paper.

%\appendix

%\section[]{}

\bsp

\label{lastpage}

\end{document}